\documentclass[amsmath,amssymb,aps,prl,reprint,groupedaddress,showpacs,letter]{revtex4-1}
\usepackage[latin9]{inputenc}
\usepackage{amsmath}
\usepackage{graphicx}
\usepackage{esint}
\usepackage[verbose]{placeins}
\usepackage{filecontents}

\makeatletter

\providecommand{\tabularnewline}{\\}

\let\textquotedbl="

\makeatother

\begin{document}

\title{Attosecond streaking delays in multi-electron systems}

\author{Tomáš Zimmermann, Lisa Ortmann, Cornelia Hofmann,
Jan-Michael Rost, Alexandra S. Landsman}
\email[]{landsman@pks.mpg.de}

\affiliation{Max Planck Institute for the Physics of Complex Systems, N\"othnitzer Stra{\ss}e 38, D-01187 Dresden, Germany}

\begin{abstract}
The use of semiclassical models based on the Strong Field Approximation (SFA) is ubiquitous in strong field multi-photon ionization and underlies many key developments in attosecond science, including the description of High Harmonic Generation (HHG).  However, such models are notably lacking in streaking experiments, which use an attosecond pulse to initiate single-photon ionization and a lower frequency infrared pulse to provide timing information.  Here, we introduce a classical Wigner propagation (CWP) method, which analogously to semiclassical models in strong field ionization, treats the ionization step quantum mechanically, followed subsequently by classical propagation (with initial conditions obtained from the Wigner function) in the infrared probe field.  As we demonstrate, this method compares well with experimental data and full two-electron quantum simulations available for helium, includes multi-electron effects, and can be applied to molecules, where the full solution of the time-dependent Schrodinger equation is not feasible.  Applying the CWP method to a many-atom molecule, like 2,3,3-trimethyl-butyl-2-iodide, we find a relatively significant directional dependence of streaking delays, indicating the importance of orientation-resolved measurements in molecules.  
\end{abstract}
\maketitle


Strong field multi-photon ionization with an infrared laser pulse underlies a number of key processes in attosecond science, including high harmonic generation (HHG) \cite{Lewenstein94,Corkum93,Schafer93}, which is behind the creation of attosecond pulses \citep{Krausz2009}.  Much of the physical interpretation of strong field ionization relies on semiclassical models based on the Strong Field Approximation (SFA) \cite{Keldysh65,PPT66,ADK86}.  These semiclassical models can reproduce experimentally measured electron momenta distributions following ionization remarkably well by combining the approximate solution to the time-dependent Schrodinger equation (TDSE) with classical propagation.  

Developing such semiclassical models for streaking experiments, which involve single photon ionization with a weak but high frequency attosecond pulse \cite{Schultze2010,Pazourek2015,Maquet2012,Calegari2016,Ossiander2017}, promises to yield much of the same benefits as it has for strong field ionization.  In particular, relative to more accurate but computationally expensive numerical solutions of the time-dependent Schrodinger equation (TDSE), semiclassical models (i) offer deeper physical insight into the ionization process;  (ii)  allow for inclusion of collisions following ionization, which can be important for denser media, such as liquids;  (iii) have in general less computational cost, allowing for modelling of more complex systems, including many-atom molecules where the solution of a TDSE becomes unfeasible.   This last point is becoming increasingly important as the field of attosecond spectroscopy, which until recently was dominated by experiments performed on rare gas atoms and metal surfaces  \citep{Gallmann2012,Peng20151}, is moving increasingly towards molecules.   

Here we develop a semiclassical framework to simulate attosecond streaking experiments:  a pump-probe scheme combining a weak attosecond XUV pump pulse, moderately strong IR probe, and the photo-electron momentum
detection \citep{Krausz2009,Schultze2010,Pazourek2015,Maquet2012,Calegari2016,Ossiander2017,Gallmann2012,Peng20151,Calegari2016}.  The principal quantity extracted from the spectrogram is the delay of the streaking trace (energy of ionized electrons plotted as a function of pump-probe delay) with respect to a reference streaking trace, which is commonly interpreted as the relative ionization delay \cite{Schultze2010,Pazourek2015}.   

To simulate the streaking traces, we introduce a Classical Wigner Propagation (CWP) method, which combines first-order perturbation theory with the classical
Wigner method \citep{Heller1976,Miller2001}. While the first-order
perturbation theory is used to describe (essentially single-photon)
interaction with the weak XUV pulse, the classical Wigner method is
used to propagate the electronic wave-packet after the ionization.
The rationale behind the latter is that semiclassical methods (exact
for linear and harmonic potentials) in general tend to perform well
for wave-packets with higher energy which are closer to the classical
limit.  

Our method has two key advantages over the existing purely classical approaches, where the binding state is represented classically by a microcanonical ensemble \citep{Pazourek2015,Nagele2011}).  In particular, we (i) treat the ionization step quantum mechanically (using perturbation theory) and (ii) describe the excited wavepacket by Wigner quasi-probability distribution, which is not only a proper phase-space density distribution with well defined classical limit of the dynamics but\textendash even when propagated
classicaly\textendash it also incorporates (some) quantum effects
through interference of positive and negative parts.

The single photon ionization process initiated by a weak attosecond XUV pulse is well-described within the first order perturbation theory.  
The wave function $\psi\left(t\right)$ is then written as \begin{widetext}
\begin{equation}
\psi\left(t\right)=\psi^{\left(0\right)}\left(t\right)+\psi^{\left(1\right)}\left(t\right)=U_{0}(t-t_{0})\psi^{\left(0\right)}\left(t_{0}\right)-\frac{i}{\hbar}\int_{t_{0}}^{t}U_{0}(t-t')H_{\text{XUV}}U_{0}(t'-t_{0})\psi^{\left(0\right)}\left(t_{0}\right)dt',\label{eq:psi_1st_order_PT}
\end{equation}
\end{widetext}where $H_{\text{XUV}}\left(t\right)=-\vec{\mu}_{e}\vec{E}_{\text{XUV}}\left(t\right)$
is the Hamiltonian of the interaction with the pump pulse, $\vec{\mu}_{e}=e\vec{q}$ is the dipole operator, $\vec{E}_{\text{XUV}}\left(t\right)=\vec{\epsilon}E_{\text{env}}\left(t\right)\cos\left(\omega_{c}t\right)$
is the pump laser field, and $U_{0}\left(t\right)=e^{\frac{-iH_{0}\left(t\right)}{\hbar}}$
is the propagator of the unperturbed Hamiltonian $H_{0}$. The unperturbed
Hamiltonian may be written as $H_{0}\left(t\right)=H_{\text{M}}+H_{\text{IR}}\left(t\right),$
where $H_{\text{IR}}\left(t\right)=\vec{\mu}_{e}\vec{E}_{\text{IR}}\left(t\right)$
is the Hamiltonian of the interaction with the IR field and $H_{\text{M}}$
is the effective single-electron Hamiltonian of the molecule. The
photo-electron momentum spectrum is computed as
\begin{equation}
S(p)=\psi_{\text{D}}^{\left(1\right)*}\left(p,t\right)\psi_{\text{D}}^{\left(1\right)}\left(p,t\right),\label{eq:spectrum}
\end{equation}
where $\psi_{\text{D}}^{\left(1\right)}\left(p,t\right)=\left\langle p\right|P_{\text{D}}\left|\psi^{\left(1\right)}\left(t\right)\right\rangle $
and $t$ is sufficiently large so that both pulses have finished and
the dissociative part of the wave packet has clearly separated from
the bound part. The projector $P_{\text{D}}$ projects out the bound
part of the wave packet. In practice, $P_{\text{D}}$ may be realized
in the position representation by setting the wave function under
some distance from the center of the molecule to zero. In addition,
only a subset of the electron momenta (typically in the direction
of the detector) may be selected by $P_{\text{D}}$ after transforming
into the momentum representation.

Evaluating the integral in Eq.~\eqref{eq:psi_1st_order_PT}, one
first has to propagate $\psi^{\left(0\right)}\left(t_{0}\right)$
with $U_{0}(t'-t_{0})$. Apart from molecular Hamiltonian $H_{\text{M}}$,
the Hamiltonian $H_{0}$ contains the interaction with the IR probe
field $H_{\text{IR}}$. Since the probe field in attosecond experiments
is typically weak enough so it does not ionize the molecule by itself
and its frequency is sufficiently low in comparison with the energy
gap separating the ground state from excited states, one may assume
that the ground state $\psi^{\left(0\right)}\left(t'\right)$ follows
$H_{0}\left(t'\right)$ adiabatically (ie., that $\psi^{\left(0\right)}\left(t'\right)$
is an eigenstate of $H_{0}$ at every time $t'$). Alternatively,
the effect of $H_{\text{IR}}$ on the ground state may be neglected
altogether. In both cases, $\psi^{\left(0\right)}\left(t'\right)$
may be computed easily by multiplying the time-independent solution
(obtained with one of the well established methods of quantum chemistry)
with the time-dependent phase factor.

After the propagation with $H_{0}\left(t'\right),$ the wave-function
$\psi^{\left(0\right)}\left(t'\right)$ is acted upon with $H_{\text{XUV}}\left(t'\right)$
resulting in the contribution to the ionizing wave packet $\psi_{\text{XUV}}\left(t'\right)=H_{\text{XUV}}\left(t'\right)\psi^{\left(0\right)}\left(t'\right)$. In order to make the subsequent approximate propagation independent of the choice of the origin of coordinates $\psi^{\left(0\right)}\left(t'\right)$ is projected out from $\psi_{\text{XUV}}\left(t'\right)$.
In contrast to $\psi^{\left(0\right)}\left(t'\right)$, subsequent
propagation of $\psi_{\text{XUV}}\left(t'\right)$ with $H_{0}\left(t'\right)$
cannot be considered adiabatic. Exploiting the fact, that in the typical
streaking experiment the dissociating state has a relatively high
kinetic energy and is generally closer to the classical limit, we
propagate $\psi_{\text{XUV}}\left(t'\right)$ with the classical Wigner
method.  (Note that in prior work \cite{Rost95}, reflection principle \cite{Heller78} was used to obtain a semi-classical approximation for total photoionization cross-section.) 

In order to apply the classical Wigner method, $\psi^{\left(1\right)}\left(t\right)$
which may now be written as 
\begin{equation}
\psi^{\left(1\right)}\left(t\right)=-\frac{i}{\hbar}\int_{t_{0}}^{t}U_{0}(t-t')\psi_{\text{XUV}}\left(t'\right)dt',\label{eq:H_XUV_psi_1}
\end{equation}
has to be transformed into the Wigner phase-space representation 
\begin{equation}
W^{\left(2\right)}(q,p,t):=\int d^{D}\xi\,\psi^{\left(1\right)}\left(q+\xi/2,t\right)\psi^{\left(1\right)*}\left(q-\xi/2,t\right)\,e^{i\xi\cdot p/\hbar}.\label{eq:Wigner_transform}
\end{equation}
Note that being quadratic in $\psi^{\left(1\right)}$, $W^{\left(2\right)}$
actually corresponds to the second-order contribution in the perturbation
theory.  Substituting the integral~\eqref{eq:H_XUV_psi_1} into Eq.~\eqref{eq:Wigner_transform}
we arrive to\begin{widetext}
\begin{equation}
W^{\left(2\right)}(q,p,t)=\frac{1}{\hbar^{2}}\int_{t_{0}}^{t}dt'\int_{t_{0}}^{t}dt''\int d^{D}\xi\,U_{0}(t-t')\psi_{\text{XUV}}\left(q+\xi/2,t'\right)\psi_{\text{XUV}}^{*}\left(q-\xi/2,t''\right)U_{0}^{\dagger}(t-t'')\,e^{i\xi\cdot p/\hbar}.\label{eq:Wigner_double_integral}
\end{equation}
\end{widetext}As will be shown in our subsequent work, this integral
may be evaluated directly within the classical Wigner method. In this
work we approximate the integral further, realizing that ``off-diagonal''
terms corresponding to different times $t'$ and $t''$ are\textendash together
with the fast oscillating carrier wave of the pump-pulse\textendash responsible
mainly for resolution of the energy spectrum of the pump pulse. Replacing the oscillating pump pulse field
$\vec{E}_{\text{XUV}}\left(t\right)$ with its envelope $\vec{E}_{\text{XUVe}}\left(t\right)=\vec{\epsilon}E_{\text{env}}\left(t\right)$,
neglecting the ``off-diagonal'' terms and taking the propagators
$U_{0}$ out of the innermost integral in Eq.~\ref{eq:Wigner_double_integral} (see Appendix for detail),
we may write the Wigner transform of $\psi_{\text{XUVe}}\left(t'\right)$
as\begin{widetext}

\begin{equation}
W_{\text{XUVe}}\left(q,p,t'\right)=\tilde{E}_{\text{env}}^{2}\left(H_{0}\left(q,p\right)-E^{\left(0\right)}-E_{c}\right)\int d^{D}\xi\,\psi_{\text{XUVe}}\left(q+\xi/2,t'\right)\psi_{\text{XUVe}}^{*}\left(q-\xi/2,t'\right)\,e^{i\xi\cdot p/\hbar},\label{eq:Wigner_transform_mu_psi}
\end{equation}
\end{widetext}where $\tilde{E}_{\text{env}}^{2}\left(E\right)$ is
the Fourier transform of the pump-pulse envelope, $H_{0}\left(q,p\right)$
Wigner transform of $H_{0}$, $E^{\left(0\right)}$ energy of $\psi^{\left(0\right)}$,
and $E_{c}=h\omega_{c}$. The propagators $U_{0}$ can be taken out
of the integral~\eqref{eq:Wigner_transform_mu_psi} due to the fact
that the propagation $U_{0}(t-t')\rho U_{0}^{\dagger}(t-t')$ which
is in the position space realized with the von~Neumann-Liouville
equation may be as well realized in the phase space with the Moyal
equation. In order to use the classical Wigner method, the Moyal equation
is simply replaced with the classical Liouville equation. Designating
the resulting classical Wigner propagation from $t'$ to $t$ with
the formal operator $U_{0}^{\text{clW}}(t-t')$ we arrive to the final
expression for $W^{\left(2\right)}(q,p,t)$
\begin{equation}
W^{\left(2\right)}(q,p,t)=\frac{1}{\hbar^{2}}\int_{t_{0}}^{t}dt'\,U_{0}^{\text{clW}}(t-t')W_{\text{XUVe}}\left(q,p,t'\right),\label{eq:Wigner_single_integral}
\end{equation}
from which the momentum spectrum is computed as
\begin{equation}
S(p)=\int d^{D}qP^{\text{W}}\left(q,p\right)W^{\left(2\right)}(q,p,t),\label{eq:Wigner_momentum_spectrum}
\end{equation}
where $P^{\text{W}}\left(q,p\right)$ is the phase-space projector
which projects out bound states localized close to the molecule and
momenta outside of the detector range.

\begin{figure}[h]
\includegraphics[width=1\columnwidth]{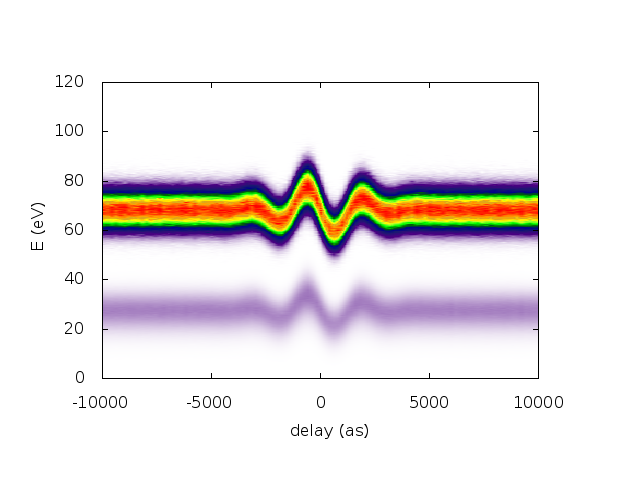}

\caption{Helium spectrogram with both shake-down and shake-up channels computed
with the CWP method at 96 eV. \label{fig:He_spectrogram}}
\end{figure}

\begin{figure}[h]
\includegraphics[width=0.8\columnwidth]{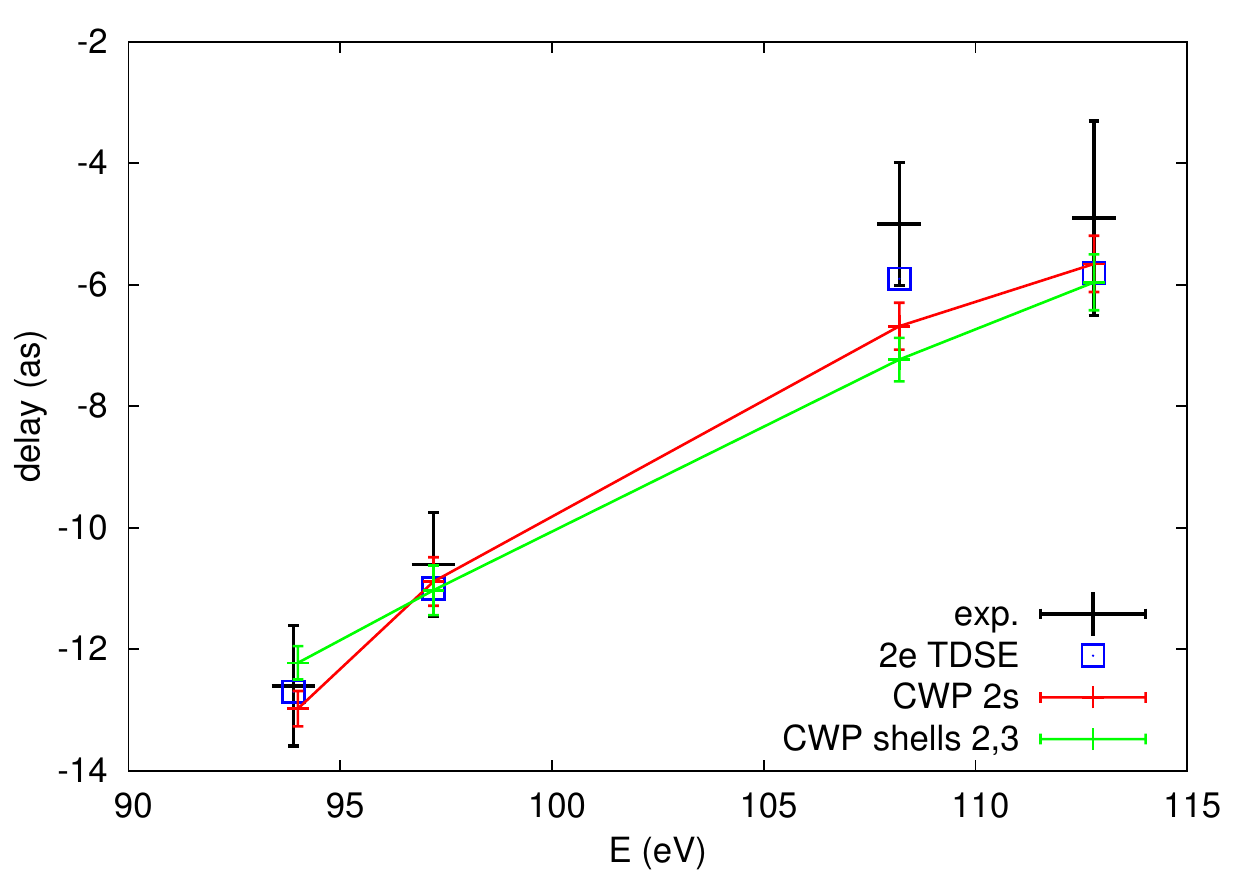}

\caption{Relative streaking delays of helium shake-down and shake-up channels. The horizontal axis shows photon energy at the center of the XUV pulse.\label{fig:He_1s_2s_delays}}
\end{figure}

\begin{figure}[h]
\includegraphics[width=0.8\columnwidth]{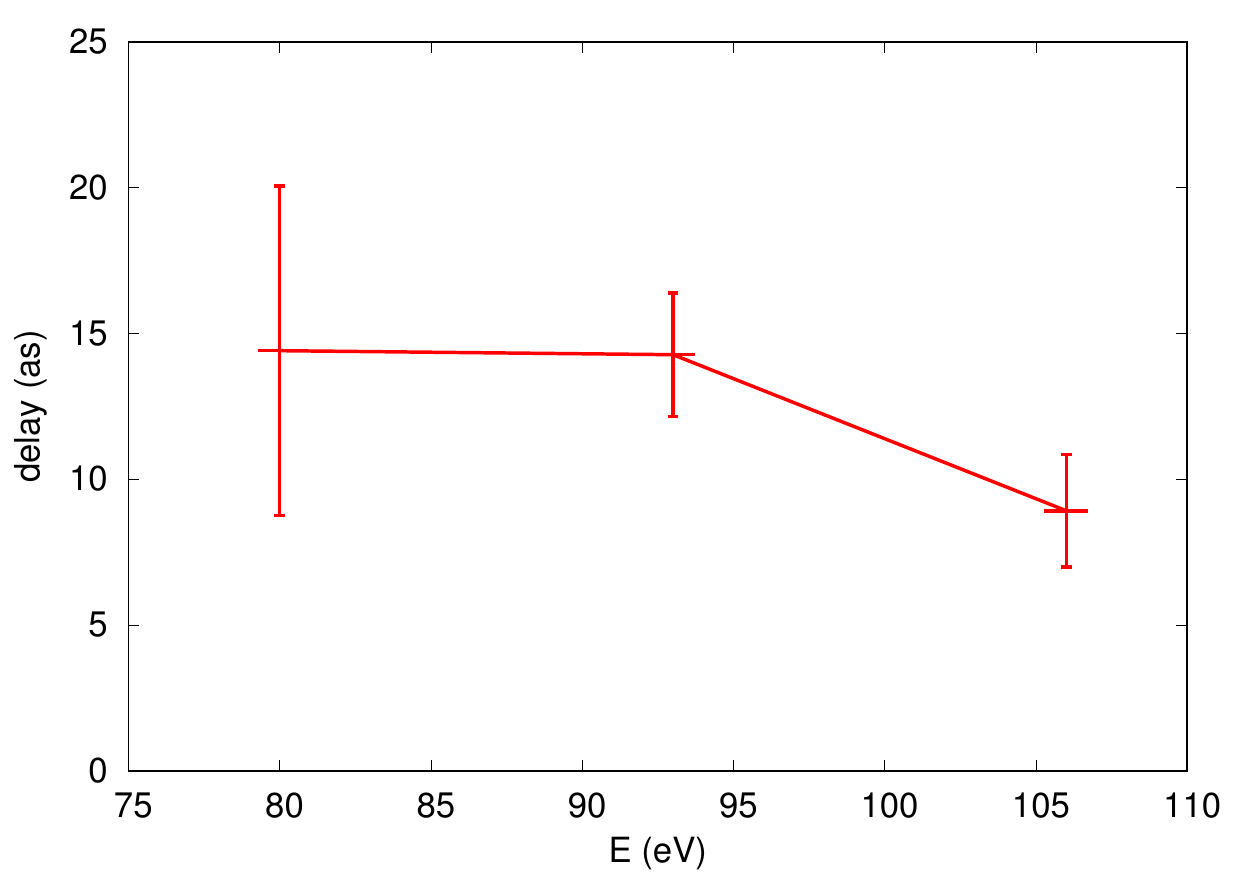}

\caption{Absolute streaking delays of 5CH$_3$EtI. The horizontal axis shows photon energy at the center of the XUV pulse.\label{fig:5CH3EtI_delays}}
\end{figure}




\begin{figure*}
\begin{tabular}{lc}
\includegraphics[width=0.7\columnwidth]{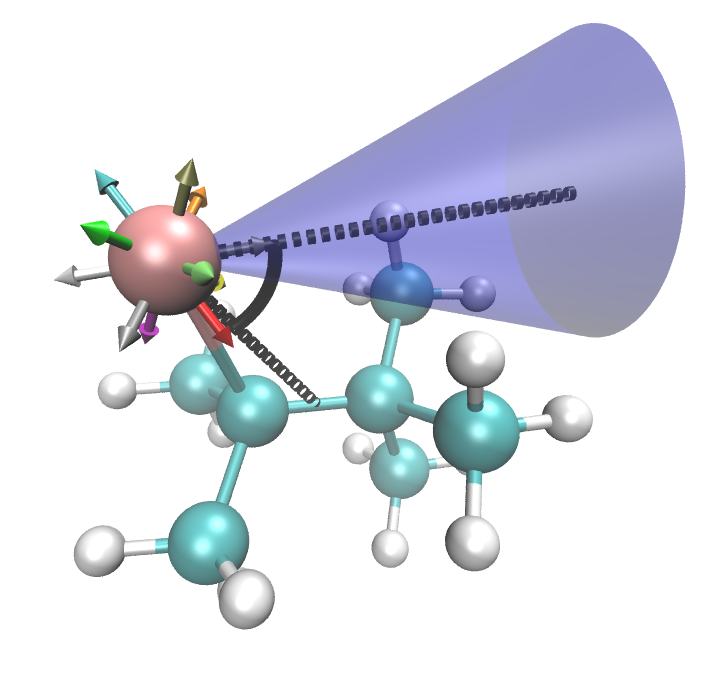} & \includegraphics[width=1\columnwidth]{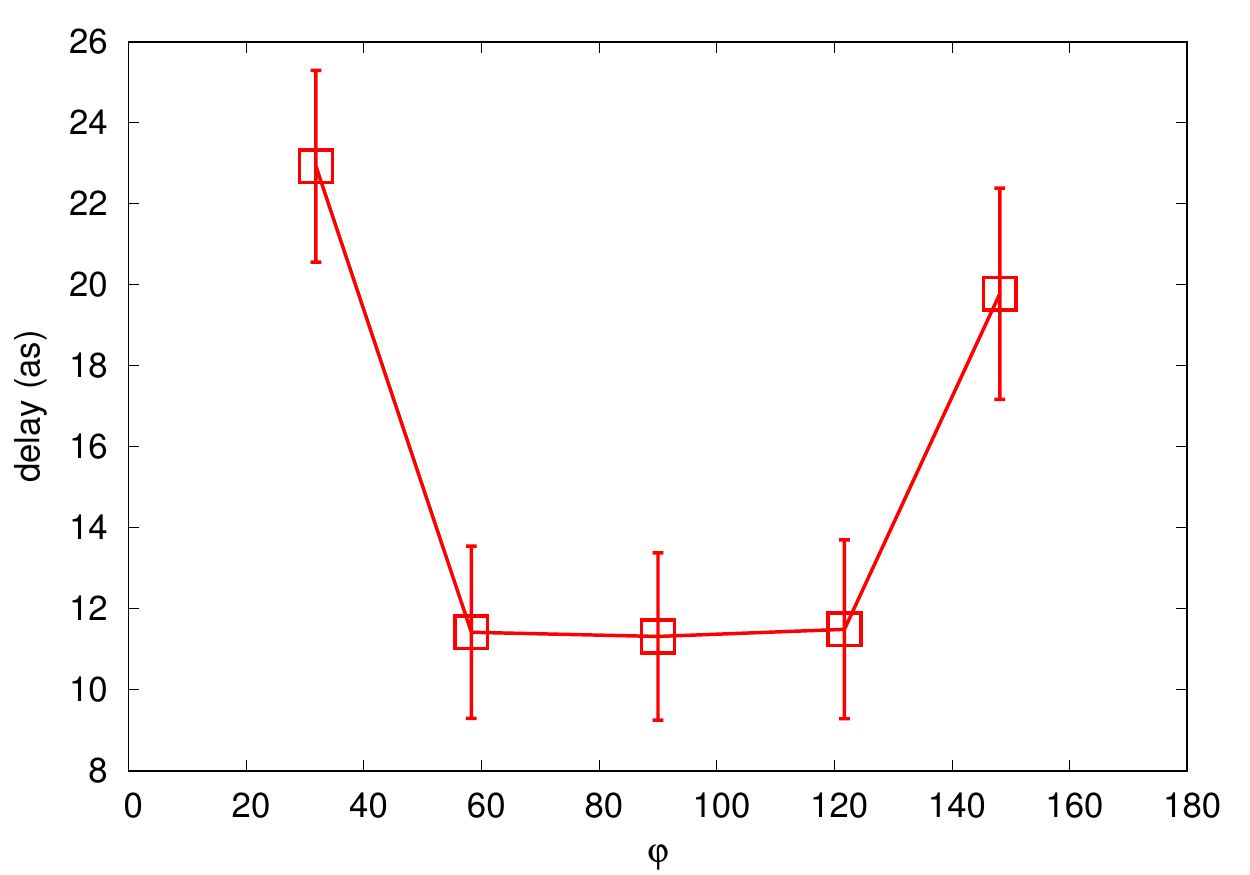}\tabularnewline
\end{tabular}

\caption{Left: 5CH$_3$EtI with 12 orientations of the polarization of the
field used for orientation averaging (arrows). The angle $\varphi$ with the
line connecting I and the center of the adjacent C-C bond is shown for one
of the orientations (in black). The photoelectron detection cone (shown
in dark blue) has a width of $20^{\circ}$ (measured from the center)
and is always parallel to the polarization of the field. Right: Orientation-dependent
stereo streaking delays in 5CH$_3$EtI iodide at 93 eV. All data points correspond to the average over two orientations with very similar $\varphi$ except for the central point at $90^{\circ}$ which is averaged over four orientations.
\label{fig:5CH3EtI_delays_orientations}}
\end{figure*}

We first apply the above expression to ionization of helium to calculate both direct and shake-up channels.  
Helium was chosen both because it is the simplest `multi-electron' system and, importantly, due to the availability of experimental data, combined with fully quantum 2-electron calculations, allowing us to benchmark the accuracy of the CWP method with a high level of certainty.  
In particular, a recent experiment has measured relative streaking delays between
the shake-down and shake-up states of helium with sub-attosecond precision
\citep{Ossiander2017}. In addition to experimental delays, Ref.~\onlinecite{Ossiander2017}
also presents delays computed with the two-electron TDSE which agrees
with the experimental values within the error bars.

The streaking trace computed using the CWP method is shown in Figure 1.  Gaussian pulse shapes with no chirp were used for both XUV and IR fields using the pulse widths and wavelengths given in Ref.~\onlinecite{Ossiander2017}. 
The initial single-electron orbitals were obtained in the Gaussian basis set as Dyson orbitals between the
neutral helium and corresponding ionic states (see Appendix for details of the Wigner transformation in the Gaussian basis set). The shake-up channel
of helium was composed from Dyson orbitals corresponding to the complete
n=2 and n=3 shells of He\textsuperscript{+}. 
This shake-up channel creates an additional streaking trace at a lower energy (see Fig. 1), and is due to the fact that some of the energy of the leaving electron may be transferred to the remaining bound electron.  To capture this multi-electron effect, it is necessary to go beyond the single active electron approximation.  In practice, it means using Dyson orbitals in calculations, rather than Hartree-Fock orbitals (see Appendix for more information on the simulation methods).


The comparison of the streaking delays computed using our CWP method with
the experimental and the two-electron TDSE delays published in Ref.~\onlinecite{Ossiander2017} is shown in Fig.~\ref{fig:He_1s_2s_delays}.
The center of mass analysis of the streaking trace followed by analytic
fitting is used to extract the CWP delays from the spectrogram. Except
for a single data point at $\sim107$ eV, the CWP result agrees with
the experiment and TDSE within error bars. To demonstrate the convergence
with the number of overlapping shake-up states two separate CWP calculations
are shown: i) calculation based solely on the Dyson orbital corresponding
to 2s state of He\textsuperscript{+} and ii) calculation based on
complete n=2 and n=3 shells of He\textsuperscript{+}. The difference
between the two is smaller than 1 as at all energies and there seems
to be no systematic trend in the error of the CWP method.




To demonstrate the ability of the CWP method to address complex molecules,
we have studied the ionization of the 4d core orbital of iodine in 2,3,3-trimethyl-butyl-2-iodide (5CH$_3$EtI). The geometry of the molecule was optimized with the MP2 method describing 36 core electrons of iodine with the model core potential (MCP) \cite{Miyoshi1998,Sekiya2001} and the valence electrons (including 4d orbital of iodine) with the cc-pVTZ basis set composed of 427
primitive Gaussians. The CIS method with the same pseudopotential and basis set was used to compute the Dyson orbitals of 4d electrons. The electrostatic potential of the positively
charged ion was computed from the single-electron density matrix obtained
with CIS/cc-pVDZ. The gas-phase streaking delays were averaged over
molecular orientations sampling the sphere with 12 directions corresponding
to edges of the small stellated dodecahedron (see Fig.~\ref{fig:5CH3EtI_delays_orientations},
note that for each direction there is a corresponding anti-parallel
direction). 

In contrast to helium or hydrogen, the orientation-averaged delays are positive. In addition,
 the orientation-resolved delays might differ from the average by almost $100\%$ of the averaged value. Figure~\ref{fig:5CH3EtI_delays_orientations} shows streaking
delays $t_{s}$ as the function of the angle $\varphi$ of the
laser polarization and the line connecting the iodine with the center of the C-C bond adjacent to the I-C bond 
(The center of this C-C bond approximately marks the center of the 5CH$_3$Et group.) 
As can be seen in the figure, the electrons which ionize along the I-C bond (in both directions)
show more positive delays $t_{s}$ than electrons ionizing in directions forming less accute $\varphi$. 
 Since the 4d orbital of I is localized on the iodine atom and almost unaffected by the presence of the 5CH$_3$Et 
group, the dependence of $t_{s}$ on $\varphi$ is a direct measure
of the influence of the functional group on the delay acquired by the photoelectron
after the ionization. 

In conclusion, we introduce a semiclassical method for calculating streaking traces for multi-electron systems.  We test the accuracy of this method by simulating the streaking trace of helium for both direct and shake-up channels, finding excellent agreement with available 2-electron TDSE simulations and experimental data. We then demonstrate that the method is capable of simulating a many-atom molecule, such as 5CH$_3$EtI, where TDSE solutions beyond the single active electron approximation may become too computationally expensive. In contrast to helium, for 5CH$_3$EtI we find positive orientation-averaged streaking delays, with relatively significant directional dependence with increased delays for electrons moving along the direction connecting iodine with the center of the 5CH$_3$Et group.



\begin{acknowledgments}
We thank Dr. Serguei Patchkovskii for kindly providing us with the
code \textsf{SuperDyson} which was used to compute CIS Dyson orbitals.
\end{acknowledgments}

\bibliographystyle{apsrev4-1}

%


\onecolumngrid
\appendix
\section{Resolution of the spectrum of XUV pump pulse}

One of the quantum interference effect in the first-order perturbation
theory which cannot be neglected is the interference due to the fast
oscillating component of the XUV field. In the CWP method, the effect
is taken into account approximately based on analogy drawn from the
fully quantum mechanical solution in the eigenbasis of the molecular
Hamiltonian. Neglecting slow oscillations of the IR field, $\psi^{\left(1\right)}\left(t\right)$
may be written as
\begin{equation}
\psi^{\left(1\right)}\left(t\right)=-\frac{i}{\hbar}\int_{t_{0}}^{t}U_{\text{mol}}\left(t-t'\right)\mu_{e}\psi^{\left(0\right)}E_{\text{env}}\left(t'\right)\frac{e^{-i\omega_{c}t'}+e^{+i\omega_{c}t'}}{2}e^{-i\omega_{g}\left(t'-t_{0}\right)}dt',\label{eq:psi_1_with_pulse}
\end{equation}
where $U_{\text{mol}}\left(t\right)=e^{-iH_{\text{mol}}t/\hbar}$,
$H_{\text{mol}}$ is the Hamiltonian of the molecule in the frozen
IR field, $\omega_{g}=E_{g}/\hbar$, $E_{g}$ is the ground state
energy of the electron in the frozen IR field, $\mu_{\epsilon}=\vec{\epsilon}\cdot\vec{\mu}_{e}$,
$E_{\text{env}}\left(t'\right)$ is the time-dependent envelope of
the pulse, and $\omega_{c}$ is the frequency of the carrier wave.
Expanding $\psi_{\mu}=\mu_{e}\psi^{\left(0\right)}$ in the basis
of eigenstates $\chi_{k}$ of $H_{\text{mol}}$ as $\psi_{\mu}=\sum_{k}\psi_{\mu,k}\chi_{k}$,
 where $\psi_{\mu,k}=\left\langle \psi_{\mu}\left|\chi_{k}\right.\right\rangle $, we get
\begin{equation}
\psi^{\left(1\right)}\left(t\right)=-\frac{i}{\hbar}\sum_{k}\psi_{\mu,k}\chi_{k}\int_{t_{0}}^{t}e^{-i\omega_{k}\left(t-t^{'}\right)}E_{\text{env}}\left(t^{'}\right)\frac{e^{-i\omega_{c}t^{'}}+e^{+i\omega_{c}t^{'}}}{2}e^{-i\omega_{g}\left(t^{'}-t_{0}\right)}dt^{'},\label{eq:psi_1_eigen_basis}
\end{equation}
where $\omega_k=E_k/\hbar$ and $E_k$ is the energy of eigenstate $\chi_{k}$.
Putting the initial time $t_{0}\rightarrow-\infty$ and the final
time to $t\rightarrow\infty$ we arrive to (omitting an overall phase
factor $e^{i\omega_{g}t_{0}}$) 
\begin{equation}
\psi^{\left(1\right)}\left(t\right)=-\frac{i}{2\hbar}\sum\psi_{\mu,k}\chi_{k}e^{-i\omega_{k}t}\left[\int_{-\infty}^{\infty}E_{\text{env}}\left(t'\right)e^{i\left(\omega_{gk}-\omega_{c}\right)t'}dt'+\int_{-\infty}^{\infty}E_{\text{env}}\left(t'\right)e^{i\left(\omega_{gk}+\omega_{c}\right)t'}dt'\right],\label{eq:psi_1_limit}
\end{equation}
where $\omega_{gk}=\omega_{k}-\omega_{g}$. Using the following definition
of the backward and forward Fourier transforms: 
\begin{align}
\tilde{f}(\omega) & :=\frac{1}{2\pi}\int_{-\infty}^{\infty}f(t)\,e^{i\omega t}dt,\label{eq:Fourier}\\
f(t) & =\int_{-\infty}^{\infty}\tilde{f}(\omega)\,e^{-i\omega t}d\omega,\nonumber 
\end{align}
$\psi^{\left(1\right)}\left(t\right)$ may be written as
\begin{equation}
\psi^{\left(1\right)}\left(t\right)=-\frac{i\pi}{\hbar}\sum\psi_{\mu,k}\chi_{k}\left(t\right)\left[\tilde{E}_{\text{env}}\left(\omega_{gk}-\omega_{c}\right)+\tilde{E}_{\text{env}}\left(\omega_{gk}+\omega_{c}\right)\right].\label{eq:psi_1_fourier}
\end{equation}
 The off-resonant term $\tilde{E}_{\text{env}}\left(\omega_{gk}+\omega_{c}\right)$
is typically very small and may be neglected in the rotating-wave
approximation (RWA). The final wave-function at $t\rightarrow\infty$
may thus be written as
\begin{equation}
\psi^{\left(1\right)}\left(t\right)=-\frac{i\pi}{\hbar}\sum_{k}\psi_{\mu,k}\chi_{k}\left(t\right)\tilde{E}_{\text{env}}\left(\omega_{gk}-\omega_{c}\right).\label{eq:psi_1_fourier_final}
\end{equation}
When the infinitesimally short delta-pulse is used, the final wave-function
is
\begin{equation}
\psi^{\left(1\right)}\left(t\right)=-\frac{i\pi}{\hbar}\sum_{k}\psi_{\mu,k}\chi_{k}\left(t\right).\label{eq:psi_1_fourier_final_delta}
\end{equation}
 Comparing Eqs.~(\ref{eq:psi_1_fourier_final}) and (\ref{eq:psi_1_fourier_final_delta}),
one can see that the energy envelope of the pulse serves as the energy
filter which chooses the eigenstates with the energy in the bandwidth
of the pulse. Analogically, as an approximation in the phase-space
picture for an arbitrary pulse, we might at every time $t'$ use the
expressions obtained for the delta pulse and multiply a resulting
phase-space distribution with the energy spectrum of the XUV pulse.
This approach is the best suited for pulses with no chirp. An extension
to chirped pulses is relatively straightforward but is not pursued
here.

\section{Wigner transformation in the real-valued Gaussian basis set}

The real-valued wave-function of the single electron orbital $\psi\left(q\right)$
may expanded in the Gaussian basis set as
\begin{equation}
\psi\left(q\right)=\sum_{j=1}^{n_{G}}c_{j}G_{j}\left(q\right),\label{eq:psi_Gaussian_expansion}
\end{equation}
where 
\begin{equation}
G_{j}(q)=\gamma_{j}\prod_{l=1}^{N_{\text{D}}}\left(q_{l}-q_{j,l}^{0}\right)^{\delta_{j,l}}\exp\left[-\epsilon_{j}\left(q_{l}-q_{j,l}^{0}\right)^{2}\right],\label{eq:Gaussian_function}
\end{equation}
$\gamma_{j}$ is the prefactor (typically product of the normalization
constant and contraction coefficient), $N_{\text{D}}$ is the number
of dimensions, $q_{j}^{0}$ is the center of the Gaussian, and $\delta_{j,l}$
is the polynomial exponent in $l$-th dimension. Substituting the
expansion (\ref{eq:psi_Gaussian_expansion}) into the generalized
Wigner transformation formula
\begin{equation}
W_{ab}(q,p,t):=\int d^{D}\xi\,\psi_{a}\left(q+\xi/2\right)\psi_{b}^{*}\left(q-\xi/2\right)\,e^{i\xi\cdot p/\hbar},\label{eq:Wigner_transform_general}
\end{equation}
we arrive to
\begin{equation}
W_{ab}(q,p)=\sum_{j=1}^{n_{G}}\sum_{k=1}^{n_{G}}c_{a,j}c_{b,k}W_{G,jk}(q,p),\label{eq:Wigner_transform_expansion}
\end{equation}
where 
\begin{equation}
W_{G,jk}\left(q,p\right)=\int d^{D}\xi\,G_{j}\left(q+\xi/2\right)G_{k}\left(q-\xi/2\right)\,e^{i\xi\cdot p/\hbar}.\label{eq:Wigner_transform_basis_functions}
\end{equation}
Substituting Eq.~(\ref{eq:Gaussian_function}) we can express $W_{G,jk}$
as a product
\begin{equation}
W_{G,jk}(q,p)=\gamma_{j}\gamma_{k}\prod_{l=1}^{N_{\text{D}}}\int d\xi_{l}\,\left(q_{l}+\xi_{l}/2-q_{j,l}^{0}\right)^{\delta_{j,l}}\left(q_{l}-\xi_{l}/2-q_{k,l}^{0}\right)^{\delta_{k,l}}g_{jk,l}\left(\xi_{l},q_{l}\right)e^{i\xi\cdot p/\hbar},\label{eq:Wigner_transform_basis_functions_expanded}
\end{equation}
where $g_{jk,l}\left(\xi_{l},q_{l}\right)=\exp\left[-\epsilon_{j}\left(q_{l}+\xi_{l}/2-q_{j,l}^{0}\right)^{2}-\epsilon_{k}\left(q_{l}-\xi_{l}/2-q_{k,l}^{0}\right)^{2}\right]$.
Completing the square, $g_{jk}\left(\xi,q\right)$ may be rewritten
as
\begin{eqnarray}
g_{jk}\left(\xi,q\right) & = & \gamma_{jk}\exp\left[-\epsilon_{jk}\left(\xi/2-q_{jk}^{0}\right)^{2}\right]\label{eq:Gaussian_product}\\
\epsilon_{jk} & = & \epsilon_{j}+\epsilon_{k}\label{eq:Gaussian_product_exp}\\
\gamma_{jk} & = & \exp\left[\frac{-\epsilon_{j}\epsilon_{k}}{\epsilon_{jk}}\left(2q-q_{j}^{0}-q_{k}^{0}\right)^{2}\right]\label{eq:Gaussian_product_prefactor}\\
q_{jk}^{0} & = & \frac{\epsilon_{j}\left(q_{j}^{0}-q\right)+\epsilon_{k}\left(q-q_{k}^{0}\right)}{\epsilon_{jk}}.\label{eq:Gaussian_product_center}
\end{eqnarray}
Expanding the polynomial product and substituting $\eta=\xi/2-q_{jk}^{0}$,
$q_{j}=q+q_{jk}^{0}-q_{j}^{0}$, and $q_{k}=-q+q_{jk}^{0}+q_{k}^{0}$,
Eq.~(\ref{eq:Wigner_transform_basis_functions_expanded}) becomes
\begin{equation}
W_{G,jk}(q,p)=2\gamma_{jk}\gamma_{j}\gamma_{k}\exp\left(2ip \cdot q_{jk}^{0}/\hbar\right)\prod_{l=1}^{N_{\text{D}}}\left(-1\right)^{\delta_{k,l}}\sum_{\alpha_{j,l}=0}^{\delta_{j,l}}\sum_{\alpha_{k,l}=0}^{\delta_{k,l}}\dbinom{\delta_{j,l}}{\alpha_{j,l}}q_{j,l}^{\alpha_{j,l}}\dbinom{\delta_{k,l}}{\alpha_{k,l}}q_{k,l}^{\alpha_{k,l}}\mathfrak{G}_{l}(p_{l},\epsilon_{jk};\alpha_{jk,l}),\label{eq:Wigner_transform_final}
\end{equation}
where $\alpha_{jk,l}=\delta_{j,l}-\alpha_{j,l}+\delta_{k,l}-\alpha_{k,l}$
and
\begin{equation}
\mathfrak{G}_{l}(p_{l},\epsilon;\alpha)=\int d\eta_{l}\,\eta_{l}^{\alpha}\exp\left(-\epsilon\eta_{l}^{2}\right)\exp\left(2i\eta_{l}p_{l}/\hbar\right).\label{eq:Gaussian_fourier_integral}
\end{equation}
For given $\alpha$, $\mathfrak{G}_{l}(p_{l},\epsilon;\alpha)$ can
be computed analytically.

\section{Simulation methods}

Initial single-electron orbitals used in the CWP method are computed
in the atom-centered Gaussian basis sets. When the correlation effects
are not important, the Hartree-Fock single-electron orbitals might
be used directly as the starting point. When the electron correlation
effects are important, more advanced electronic structure theory may
be used. In that case, single-electron orbitals are obtained as Dyson
orbitals between the wave-functions of the molecule and the corresponding
ion. Often, more single-electron orbitals contribute to the spectrum
(see the following paragraph for the details of the sampling algorithm
in that case).

For each phase-space point sampled during the calculation, the Wigner
transforms were computed using analytic Gaussian integrals (for details
see above). The Metropolis algorithm was used to sample
the phase-space (using the discrete time step $\Delta t=2$ au on
the time axis). Since the Wigner function is not positive definite,
the acceptance probability is given by the absolute value of $W_{\text{XUVe}}\left(q,p,t'\right)$
and the sign of $W_{\text{XUVe}}\left(q,p,t'\right)$ is included
in the statistical weight. When more than one orbital contributes to
the streaking trace, the Metropolis algorithm is modified further:
1) the acceptance probability is given by the sum of absolute values
of probabilities of all contributing orbitals and 2) second stochastic
step is added in order to choose the orbital and therefore
the sign of the statistical weight of the trajectory. The probability for each orbital in this step is given
by the ratio of $\left|W_{\text{XUVe}}\left(q,p,t'\right)\right|$
for the orbital and the total sum of absolute values.
For each time $t'$, the trajectories sampled were propagated with
the Cash-Karp variant of the adaptive-step Runge-Kutta method.

The streaking spectra were computed using 64 to 128 time slices. The
absolute photo-ionization delays $t_{D}$ were measured with respect to
the vector potential of the IR pulse and were extracted by fitting
the streaking trace with the function $f\left(t\right)=a\exp\left[-g\left(t-h\right){}^{2}\right]\sin\left(bt+c\right)+d$
and using $t_{D}=c/b$. The streaking trace was obtained either as
a curve connecting centers of mass (COM) of the intensity profile
of each time slice (TDSE and CWP) or as a curve connecting maxima
of the intensity profile (TDSE only).

The electronic structure was computed with \textsf{Gamess} (CIS, HF)
\citep{Schmidt1993} and \textsf{QChem} (EOM-CCSD) \citep{Shao2015}.
The Dyson orbitals for correlated methods were computed with the \textsf{Superdyson}
code (CIS) \citep{Patchkovskii2006} and \textsf{QChem} (EOM-CCSD).
The Gaussian integrals needed to compute electrostatic potential and
forces of the ion were computed with the \textsf{libcint} library
\citep{Sun2015}. The analytic form of Gaussian Fourier integrals
which appear in the Wigner transformation was found using \textsf{SageMath}.

In order to compute
the relative streaking delays for two-electron helium, the neutral, shake-down,
and shake-up states were described with EOM-CCSD theory (exact for
two electron systems) in the custom Gaussian basis set necessary to
properly describe n=3 shell of He\textsuperscript{+}. The initial
single-electron orbitals were obtained as Dyson orbitals between the
neutral helium and corresponding ionic states. The shake-up channel
of helium was composed from Dyson orbitals corresponding to the complete
n=2 and n=3 shells of He\textsuperscript{+}. The first-order multipole
expansion was used to represent the potential of the helium ion.

\bibliographystyle{apsrev}

\end{document}